\documentclass[aps,nofootinbib,prd,eqsecnum,showpacs,showkeys,preprintnumbers]{revtex4-2}
\usepackage[caption=false]{subfig}
\usepackage[utf8]{inputenc}
\usepackage{graphicx}
\usepackage{amsmath}
\usepackage{amsfonts}
\usepackage{amssymb}
\usepackage{float}
\usepackage{color}
\usepackage{bm}
\usepackage{mathrsfs}
\usepackage{epstopdf}
\usepackage{url}
\usepackage{footnote}
\usepackage{textcomp}
\usepackage{ulem}
\usepackage{esint}
\usepackage[unicode=true, pdfusetitle,
 bookmarks=true,bookmarksnumbered=false,bookmarksopen=false,
 breaklinks=false,pdfborder={0 0 1},backref=false,colorlinks=false]{hyperref}
\usepackage{multirow}
\usepackage{pifont}
\usepackage{soul}
\begin{document}

\title{Insights into Gravitinos Abundance, Cosmic Strings and Stochastic
Gravitational Wave Background}
\author{K. El Bourakadi$^{1,2}$}
\email{k.elbourakadi@yahoo.com}
\author{G. Otalora$^{1}$}
\email{giovanni.otalora@academicos.uta.cl}
\author{A. Burton-Villalobos$^{1}$}
\email{andres.burton.villalobos@alumnos.uta.cl}
\author{H. Chakir$^{2}$}
\email{chakir10@gmail.com }
\author{M. Ferricha-Alami$^{3}$}
\email{ferrichaalami@yahoo.fr}
\author{M. Bennai$^{3}$}
\email{mdbennai@yahoo.fr}

\begin{abstract}
In this paper, we investigate D-term inflation within the framework of
supergravity, employing the minimal K\"{a}hler potential. Following previous studies that revealed
that this model can overcome the $\eta$-problem found in F-term models, we explore reheating dynamics and gravitino production, emphasizing the interplay between reheating temperature, spectral index, and
gravitino abundance. Our analysis indicates that gravitino production is
sensitive to the equation of state during reheating, affecting the reheating
temperature and subsequent dark matter relic density. Furthermore, we analyze gravitational waves
generated by cosmic strings, providing critical constraints on early
Universe dynamics and cosmic string properties, the energy scales of both inflation and string formation influence the stochastic gravitational wave background (SGWB) generated by these cosmic strings.
\end{abstract}

\affiliation{$^{1}$Departamento de F\'isica, Facultad de Ciencias,
Universidad de Tarapac\'a, Casilla 7-D, Arica, Chile.\\} 
\affiliation{$^{2}$\small Subatomic Research and Applications Team, Faculty of Science
Ben M'sik,\\
{\small Casablanca Hassan II University, Morocco}\\} 
\affiliation{$^{3}${\small Quantum Physics and Magnetism Team, LPMC, Faculty of Science
Ben M'sik,}\\
{\small Casablanca Hassan II University, Morocco.}\\}

\maketitle

\section{Introduction}

Supersymmetric (SUSY) inflation \cite{A1,A2,A3,A4} provides an appealing
framework for connecting inflation with particle physics models rooted in
grand unified theories (GUTs) \cite{A5}. In the simplest models, the soft
supersymmetry breaking terms are crucial for aligning predictions of the
scalar spectral index with cosmic microwave background (CMB) observations 
\cite{A6},\cite{A7}. SUSY provides an appealing solution to the analogous
hierarchy problem in the Standard Model (SM) of particle physics and also
facilitates the unification of the three gauge couplings. Notably, the local
version of supersymmetry --- supergravity (SUGRA) --- would dominate the
dynamics of the early Universe when high-energy physics played a crucial
role. Therefore, considering inflation within the framework of supergravity
is quite natural. However, integrating inflation into supergravity is a
challenging task. This difficulty arises mainly because the SUSY breaking
potential term, essential for inflation, typically imparts an additional
mass to the would-be inflaton, thereby disrupting the flatness of the
inflaton potential \cite{A8,A9}. To achieve successful inflation that aligns
with observational data from large-scale structures and the anisotropy of
CMB radiation, the potential of the scalar field, known as the inflaton,
must be very flat. This required flatness of the potential can be achieved
through the mechanisms provided by supersymmetry or supergravity. In
supersymmetric models, the scalar potential is composed of contributions
from both the F-term and D-term.

Specifically, in F-term inflation models, where the vacuum energy that
drives the inflationary expansion predominantly comes from the F-term, the
inflaton mass is generally on the same order as the Hubble parameter $H$
during inflation \cite{A10}. Consequently, achieving a sufficiently long
expansion to address the horizon and flatness problems is challenging. This
issue is referred to as the eta problem in inflation models within the
framework of supergravity. Conversely, D-term inflation, in which the vacuum
energy is supplied by the D-term, does not encounter this problem \cite%
{A11,A12}. Therefore, from this perspective, D-term inflation is more
appealing than F-term inflation. However, it has been discovered that the
D-term inflation model also has its own set of issues \cite{A13}. For
instance, from an observational standpoint, cosmic strings generated after
inflation can significantly impact the spectrum of CMB anisotropy \cite{A14}%
, as this model is a type of hybrid inflation. Additionally, there are
concerns about the validity of the D-term inflation potential, since the
inflaton requires a large initial value on the order of the (sub-)Planck
scale for a natural model parameter\cite{A15}. Therefore, D-term inflation
appears to be under considerable scrutiny. Additionally, within the
framework of (heterotic) string models, there are two further issues: the
runaway behavior of the dilaton and an excessively large magnitude of the
Fayet-Iliopoulos (FI) term \cite{A16}.

Given the current and anticipated future constraints on inflaton decay via
CMB limits on $N_{re}$ and the competition with constraints from gravitino
abundance in supersymmetric models, we revisit the issue of gravitino
production following inflation. A well-understood source of gravitinos is
their production through particle collisions in the thermal plasma that
fills the Universe after reheating \cite{A17,A18,A19}. However, gravitinos
could also have been produced by particle collisions before the reheating
process was complete, either from collisions of relativistic inflaton decay
products before thermalization or within any dilute thermal plasma formed by
these collisions while inflaton decay was still ongoing \cite{A19-1} Cosmic
strings are intriguing messengers from the early Universe due to their
distinctive signatures in the stochastic gravitational wave background
(SGWB). Recent NANOGrav 12.5-year data has provided evidence of a stochastic
process at nanohertz frequencies, which has been interpreted as SGWB in
numerous recent studies \cite{A20,A21,A22,A23}. Relic gravitational waves
(GWs) offer a fascinating window into the exploration of early Universe
cosmology \cite{A24}. Cosmic strings generate powerful bursts of
gravitational radiation that could be detected by interferometric
gravitational wave detectors such as LIGO, Virgo, and LISA \cite{A25,A26}.
Additionally, the SGWB can be detected or constrained through various
observations, including big bang nucleosynthesis (BBN), pulsar timing
experiments, and interferometric gravitational wave detectors \cite{A27}.
The study of scalar induced gravitational waves offers crucial insights into
the early Universe, particularly within the inflationary paradigm. Detecting
the primordial gravitational wave background can substantiate inflationary
models by providing evidence for a kinetic epoch preceding inflation,
characterized by a distinct blue tilt at higher frequencies. The
perturbation theory framework reveals that the transverse-traceless
components of these waves evolve according to second-order gravitational
interactions induced by first-order curvature perturbations. The spectral
energy density of these waves at various scales, particularly those relevant
to Pulsar Timing Array (PTA) observations, can be quantified through
analytical models, which align well with empirical data, such as that from
NANOGrav. These models illustrate the dependence of gravitational wave
amplitudes on key cosmological parameters, thus highlighting the interplay
between inflationary dynamics and observable gravitational wave spectra. The
analysis of gravitational waves generated by cosmic strings provides
critical constraints on the dynamics of the early Universe and the
properties of cosmic strings. The dimensionless string tension, $G\mu _{cs}$%
, significantly influences the stochastic gravitational wave background
(SGWB) produced by cosmic strings post-inflation. The SGWB from cosmic
strings is shaped by the energy scales of inflation and the string
formation, with the amplitude of tensor mode anisotropy fixing the inflation
scale around $\Lambda _{\inf }\sim 3.3\times 10^{16}r^{1/4}$. Cosmic string
contributions to the gravitational wave spectrum, represented by $%
\Omega_{GW}(f)$, depend on the number density and dynamics of cosmic string
loops across different cosmological epochs. These contributions exhibit a
strong frequency dependence and vary significantly with the string tension
parameter, highlighting the complex relationship between cosmic string
properties and observable gravitational wave signals. The comprehensive
models and visualizations underscore the importance of cosmic string
dynamics in shaping the SGWB and provide insights into potential
observational signatures in gravitational wave experiments, aiding in the
broader understanding of early universe cosmology.

The paper is organized as follows, in Sec. \ref{sec1} we examine
supergravity formalism, in Sec. \ref{sec2} we focus on the D-term model for
the case of the minimal K\"{a}hler potential. In Sec. \ref{sec3}, we apply
the previous findings in the study of the reheating process and its relation
to gravitinos production. In Sec. \ref{sec4}, we link the primordial
gravitational waves production from different epochs in the light of the
chosen supergravity formalism, which is constrained by PTA observations. In
Sec. \ref{sec5}, we study the gravitational waves production due to networks
of cosmic strings. In Sec. \ref{sec6}, we conclude.

\section{Supergravity Formalism}

\label{sec1}

The supergravity lagrangian scalar part can be defined by the superpotential 
$W\left( \Theta _{i}\right) $, K\"{a}hler potential $K\left( \Theta
_{i},\Theta _{i}^{\ast }\right) $, and gauge kinetic function $f\left(
\Theta _{i}\right) $ \cite{B1,B2,B3}. $W$ and $f$ are holomorphic functions
of complex scalar fields, whereas the initial function $K$ is
non-holomorphic and operates as a real function of the scalar fields $\Theta
_{i}$ and their conjugates $\Theta _{i}^{\ast }$. The three functions
mentioned above are initially defined in terms of (chiral) superfields. Since our primary focus is on the scalar component of a superfield, we equate a
superfield with its complex scalar counterpart, denoting both by the same
symbol. The action of the complex scalar fields, when minimally coupled to
gravity, comprises both kinetic and potential components.%
\begin{equation}
S=\int d^{4}x\sqrt{-g}\left[ \frac{1}{\sqrt{-g}}{\mathcal{L}}_{kin}-V\left(
\Theta _{i},\Theta _{i}^{\ast }\right) \right] .
\end{equation}

The kinetic terms of the scalar fields are dictated by the K\"{a}hler
potential, denoted as K, and expressed as follows:%
\begin{equation}
\frac{1}{\sqrt{-g}}{\mathcal{L}}_{kin}=-K_{ij\ast }D_{\mu }\Theta _{i}D_{\nu
}\Theta _{j}^{\ast }g^{\mu \nu },
\end{equation}%
with $K_{ij\ast }=\partial ^{2}K/\partial \Theta _{i}\partial \Theta
_{j}^{\ast },$\ and $D_{\mu }$ denotes the covariant derivative in the gauge
field context. The potential $V$ for scalar fields $\Theta _{i}$ consists of
two components, namely, the F-term $V_{F}$ and the D-term $V_{D}$. Here the
F-term component which is determined by both the superpotential $W$ and the K%
\"{a}hler potential $K$ is given by,%
\begin{equation}
V_{F}=e^{K}\left[ D_{\Theta _{i}}WK_{ij\ast }^{-1}D_{\Theta _{j}^{\ast
}}W^{\ast }-\frac{3\left\vert W\right\vert ^{2}}{M_{p}^{2}}\right] ,
\label{eq2-3}
\end{equation}%
where

\begin{equation}
D_{\Theta _{i}}W=\frac{\partial W}{\partial \Theta _{i}}+\frac{1}{M_{p}^{2}}%
\frac{\partial K}{\partial \Theta _{i}}W.
\end{equation}%
Conversely, the D-term $V_{D}$ is associated with gauge symmetry and is
defined by the gauge kinetic function and the K\"{a}hler potential,%
\begin{equation}
V_{D}=\frac{1}{2}\sum \left[ Ref_{ab}\left( \Theta _{i}\right) \right]
^{-1}g_{a}g_{b}D_{a}D_{b},  \label{eq2-5}
\end{equation}%
where%
\begin{equation}
D_{a}=\Theta _{i}\left( T_{a}\right) _{j}^{i}\frac{\partial K}{\partial
\Theta _{i}}+\xi _{a}.
\end{equation}

In this context, the subscripts $\left( a\right) $ and $\left( b\right) $
denote gauge symmetries, where $g_{a}$ represents the gauge coupling
constant, and $T_{a}$ stands for the corresponding generator. The term $\xi
_{a}$ is referred to as the Fayet--Iliopoulos $\left( FI\right) $ term,
which is non-zero exclusively when the gauge symmetry is Abelian,
specifically $U(1)$ symmetry. Keep in mind that within a supersymmetric
framework, the tree-level potential during inflation comprises both an
F-term and a D-term. These two components exhibit distinct characteristics,
and in the context of all inflationary models, it is noteworthy that
dominance is typically attributed to only one of these terms.

In certain models rooted in Supergravity, such as the one under
consideration here with a nonminimal K\"{a}hler potential, the exponential
term $e^{K}$ contributes a term $V$ to the effective mass squared, roughly
of the order of $H^{2}$ (the Hubble scale squared), affecting all scalar
fields. Consequently, inflation experiences an effective mass squared $%
V_{g}=\sum \left\vert \frac{\partial W}{\partial \Theta _{i}}\right\vert
^{2}=3H^{2}$ \cite{BB3}, introducing a contribution to the slow-roll
parameter $\eta $. This contribution results in a violation of the slow-roll
conditions ($\left\vert \eta \right\vert \ll 1$) and leads to the so-called $%
\eta $-problem

\begin{equation*}
\eta =M_{p}^{2}\frac{V^{\prime \prime }}{V}\simeq \frac{1}{3}\left( \frac{m}{%
H}\right) ^{2}\simeq 1.
\end{equation*}

Addressing this challenge has prompted the proposal of various approaches.
Here, we are only interested in the use of D-term hybrid inflation, which can provide positive energy in D-term potential. This allows for the realization of inflation while avoiding the $\eta$-problem.

\section{D-term Hybrid Model}

\label{sec2}

Constructing an inflation model in supergravity faces a significant
challenge arising from the F-term, particularly the exponential factor
associated with it. Hence, achieving a positive potential energy in the
D-term could pave the way for successful inflation without encountering the $%
\eta $ problem. This insight was initially highlighted in Ref. \cite{B4}.

Let us consider a D-term model of hybrid inflation as introduced in \cite%
{B5,B6},%
\begin{equation}
W=\lambda S\Theta _{+}\Theta _{-},
\end{equation}

where $\lambda $ is the coupling constant, and $S,$\ $\Theta _{+},$\ and $%
\Theta _{-}$ are three (chiral) superfields. The superpotential remains
unchanged under a $U(1)$ gauge symmetry, with assigned charges of $0$, $+1$,
and $-1$ for the fields $S$, $\Theta _{+}$, and $\Theta _{-}$, respectively.
Additionally, it exhibits an \textit{R} symmetry governing the
transformation of these fields as follows:%
\begin{equation}
S\left( \theta \right) \rightarrow e^{2i\alpha }S\left( \theta e^{i\alpha
}\right) ,~~~~~~\Theta _{+}\Theta _{-}(\theta )\rightarrow \Theta _{+}\Theta
_{-}(\theta e^{i\alpha }),
\end{equation}%
and considering the minimal K\"{a}hler potential,%
\begin{equation}
K=\left\vert S\right\vert ^{2}+\left\vert \Theta _{+}\right\vert
^{2}+\left\vert \Theta _{-}\right\vert ^{2}.
\end{equation}

From Eqs. \eqref{eq2-3} and \eqref{eq2-5} the tree level scalar potential is given by,%
\begin{eqnarray}
V\left( S,\Theta _{+},\Theta _{-}\right) &=&\lambda ^{2}e^{\left\vert
S\right\vert ^{2}+\left\vert \Theta _{+}\right\vert ^{2}+\left\vert \Theta
_{-}\right\vert ^{2}}\left[ \left\vert \Theta _{+}\Theta _{-}\right\vert
^{2}+\left\vert S\Theta _{-}\right\vert ^{2}+\left\vert S\Theta
_{+}\right\vert ^{2}+\left( \left\vert S\right\vert ^{2}+\left\vert \Theta
_{+}\right\vert ^{2}+\left\vert \Theta _{-}\right\vert ^{2}+3\right)
\left\vert S\Theta _{+}\Theta _{-}\right\vert ^{2}\right]  \notag \\
&&+\frac{g^{2}}{2}\left( \left\vert \Theta _{+}\right\vert ^{2}-\left\vert
\Theta _{-}\right\vert ^{2}+\xi \right).
\end{eqnarray}

Here, $g$ represents the gauge coupling constant, $\xi $ is a non-zero FI
term, and we have adopted a minimal gauge kinetic function. This potential
exhibits a distinctive global minimum $V=0$ at $S=\Theta _{+}=0,\Theta _{-}=%
\sqrt{\xi }.$ Yet, when $\left\vert S\right\vert $ is significantly large,
the potential displays a local minimum with a positive energy density at $%
\Theta _{+}=\Theta _{-}=0$. To determine the critical value $S_{c}$ of $%
\left\vert S\right\vert $, we compute the mass matrix of $\Theta _{+}$ and $%
\Theta _{-}$ along the inflationary trajectory where $\Theta _{+}=\Theta
_{+}=0$. This is expressed as:%
\begin{equation}
V_{mass}=m_{+}^{2}\left\vert \Theta _{+}\right\vert ^{2}+m_{-}^{2}\left\vert
\Theta _{-}\right\vert ^{2},
\end{equation}%
with%
\begin{equation}
m_{+}^{2}=\lambda ^{2}\left\vert S\right\vert ^{2}e^{\left\vert S\right\vert
^{2}}+g^{2}\xi ,~~~~~~~~~m_{-}^{2}=\lambda ^{2}\left\vert S\right\vert
^{2}e^{\left\vert S\right\vert ^{2}}-g^{2}\xi .
\end{equation}

Hence, provided that $m_{-}^{2}\geq 0$, corresponding to $\left\vert
S\right\vert \geq S_{c}\simeq g\sqrt{\xi }/\lambda $ for $S_{c}\lesssim 1$,
the local minimum at $\Theta _{+}=\Theta _{+}=0$ remains stable, leading to
inflation driven by the positive potential energy density of $g^{2}\xi
^{2}/2 $. Moreover, this mass split induces quantum corrections computed
using the standard formula \cite{BB3,B7},%
\begin{equation}
V_{1L}=\frac{1}{32\pi ^{2}}\left[ m_{+}^{4}\ln \left( \frac{\lambda
^{2}\left\vert S\right\vert ^{2}e^{\left\vert S\right\vert ^{2}}+g^{2}\xi }{%
\Lambda ^{2}}\right) +m_{-}^{4}\ln \left( \frac{\lambda ^{2}\left\vert
S\right\vert ^{2}e^{\left\vert S\right\vert ^{2}}+g^{2}\xi }{\Lambda ^{2}}%
\right) -2\lambda ^{4}\left\vert S\right\vert ^{4}e^{\left\vert
2S\right\vert ^{2}}\ln \left( \frac{\lambda ^{2}\left\vert S\right\vert
^{2}e^{\left\vert S\right\vert ^{2}}}{\Lambda ^{2}}\right) \right] ,
\end{equation}%
and when $\left\vert S\right\vert \gg S_{c},$\ the effective potential of \sout{S} $S$ during inflation is approximated as,%
\begin{equation}
V\left( S\right) \simeq \frac{g^{2}\xi ^{2}}{2}\left[ 1+\frac{g^{2}}{8\pi
^{2}}\ln \left( \frac{\lambda ^{2}\left\vert S\right\vert ^{2}e^{\left\vert
S\right\vert ^{2}}}{\Lambda ^{2}}\right) \right] .
\end{equation}

Given that the potential is independent of the phase of the complex scalar
field $S$, we can propose its real part as $\sigma \equiv \sqrt{2}ReS$.
Subsequently, for $\sigma _{c}\ll \sigma \lesssim 1$, the effective
potential of the inflaton field $\sigma $ is expressed as:%
\begin{equation}
V\left( \sigma \right) \simeq \frac{g^{2}\xi ^{2}}{2}\left[ 1+\frac{g^{2}}{%
8\pi ^{2}}\ln \left( \frac{\lambda ^{2}\sigma ^{2}}{2\Lambda ^{2}}\right) %
\right] .
\end{equation}

During the inflationary period, the slow roll slow-roll parameters take the form%
\begin{eqnarray}
\epsilon &\simeq &\frac{g^{4}}{32\pi ^{4}\sigma ^{2}}, \\
\eta &\simeq &-\frac{g^{2}}{4\pi ^{2}\sigma ^{2}},
\end{eqnarray}%
and the e-folding number \textit{N} is estimated as%
\begin{equation}
N\simeq \frac{2\pi ^{2}}{g^{2}}\left( \sigma _{k}^{2}-\sigma
_{end}^{2}\right) .
\end{equation}

Throughout the inflationary regime, curvature perturbations arose from
inflaton fluctuations. The amplitude of these perturbations in the comoving
gauge ${\mathcal{R}}$ \cite{B8,Gonzalez-Espinoza:2019ajd,Gonzalez-Espinoza:2020azh,Leyva:2021fuo,Leyva:2022zhz}, measured on the comoving scale of $2\pi /k$%
, is determined as,%
\begin{equation}
{\mathcal{R}}^{2}\simeq \frac{1}{24\pi ^{2}}\frac{V}{\epsilon }=\frac{N}{3}%
\xi ^{2}.
\end{equation}

Here, $k$ represents the epoch when the $k$ mode exited the Hubble radius in
the course of inflation, as indicated in \cite{B9,B10,B11}. Conversely,
gravitational wave tensor perturbations $\mathit{h}$ are also generated, and
their amplitude on the comoving scale of $2\pi /k$ is specified in Ref. \cite{B12}%
, and calculated in this model as,%
\begin{equation}
h_{k}^{2}\simeq \frac{2}{3\pi ^{2}}V\left( \sigma _{k}\right) =\frac{%
g^{2}\xi ^{2}}{3\pi ^{2}}.
\end{equation}%
Then, the tensor-to-scalar ratio $r$ is given by 
\begin{equation}
r\equiv \frac{h_{k}^{2}}{{\mathcal{R}}^{2}}\simeq 16\epsilon
\end{equation}%
where $h_{k}^{2}=2H_{k}^{2}/\pi ^{2}$, and for a chosen pivot scale, the
power spectrum of scalar perturbation can be considered as ${\mathcal{R}}%
^{2}\simeq A_{s}$.\ 

\section{ Reheating Dynamics and Gravitinos Production}

\label{sec3}

In the initial stages, the quasi-de-Sitter phase is driven by the inflaton
field, resulting in $N_{k}$ e-folds of expansion. The comoving horizon scale
decreases proportionally to $\sim a^{-1}$ during this period. Following the
conclusion of accelerated expansion and the subsequent expansion of the
comoving horizon, the reheating phase begins \cite{Kofman:1994rk}. After an additional $N_{re}$
e-folds of expansion, all the energy stored in the inflaton field is
completely dissipated, leading to the formation of a hot plasma with a
reheating temperature of $T_{re}$. Following this phase, the Universe
undergoes $N_{RD}$ e-folds of expansion in a state of radiation domination
before transitioning into a state of matter domination.

In cosmology, we observe perturbation modes that exhibit magnitudes
comparable to the size of the cosmic horizon. For instance, Planck
identifies the pivot scale at $k=0.05Mpc^{-1}$ \cite{Ade:2015lrj}. The comoving Hubble scale,
denoted as $a_{k}H_{k}=k$ is associated with the current timescale in
relation to the moment when this particular mode crossed the horizon \cite%
{B112,B113},%
\begin{equation}
\frac{k}{a_{0}H_{0}}=\frac{a_{k}}{a_{end}}\frac{a_{end}}{a_{re}}\frac{a_{re}%
}{a_{eq}}\frac{a_{eq}H_{eq}}{a_{0}H_{0}}\frac{H_{k}}{H_{eq}}.  \label{eq14}
\end{equation}%
Quantities represented by a subscript $\left( k\right) $ are calculated at
the point of horizon exit. Other subscripts denote various epochs, including
the end of inflation $(_{end})$, reheating $(_{re})$, radiation-matter
equality $(_{eq})$, and the present time $(_{0})$. It is noteworthy that $%
\ln \left( \frac{a_{k}}{a_{end}}\right) =N_{k},$\ $\ln \left( \frac{a_{end}}{%
a_{re}}\right) =N_{re}$\ and $\ln \left( \frac{a_{re}}{a_{eq}}\right)
=N_{RD}.$

A connection between the temperature at the end of reheating $T_{re}$ and
the CMB temperature $T_{0}$ can be established by considering factors is
given by, 
\begin{equation}
T_{re}=\left( \frac{43}{11g_{re}}\right) ^{\frac{1}{3}}\left( \frac{%
a_{0}T_{0}}{k}\right) H_{k}e^{-N_{k}}e^{-N_{re}}.
\end{equation}%
On the other hand, the reheating duration can be expressed as,%
\begin{equation}
N_{re}=\frac{1}{3\left( 1+\omega _{re}\right) }\ln \left( \frac{30\cdot 
\frac{3}{2}V_{end}}{\pi ^{2}g_{re}T_{re}^{4}}\right) ,
\end{equation}%
and thus, the final form of reheating temperature is described as a function of the
number of inflationary e-foldings $N_{k}$, the Hubble parameter $H_{k}$\ and
the potential value at the end of inflation $V_{end}$\ \ \ as \cite%
{B10,B11,Lopez:2021agu},

\begin{equation}
T_{re}=\left[ \left( \frac{43}{11g_{re}}\right) ^{\frac{1}{3}}\frac{%
a_{0}T_{0}}{k}H_{k}e^{-N_{k}}\left[ \frac{3^{2}\cdot 5V_{end}}{\pi ^{2}g_{re}%
}\right] ^{-\frac{1}{3\left( 1+\omega _{re}\right) }}\right] ^{\frac{3\left(
1+\omega _{re}\right) }{3\omega _{re}-1}}.  \label{eq20}
\end{equation}%
\ \ To determine the reheating temperature, $T_{re}$, for a specific model,
one must calculate $N_{k}$, $H_{k}$, $V_{end}$, and the potential at the end of inflation by using the formula $V_{end}=V\left( \sigma
_{end}\right) $, and knowing that $\sigma _{end}$\ is determined considering $%
\left\vert \eta \right\vert =1.$\ \ 

The SUGRA effects result in the coupling of the inflaton field $\phi $ to
all matter fields, provided there is a non-zero vacuum expectation value (%
\textit{VEV}). The interactions with fermions are appropriately expressed in
the context of the total K\"{a}hler potential, denoted as $G=K+\ln |W|^{2}$, and such that:%
\begin{equation}
{\mathcal{L}}=-\frac{1}{2}e^{G/2}G_{\phi ijk}\phi \psi ^{i}\psi ^{j}\varphi
^{k}+h.c.
\end{equation}%
In the given context, $\varphi ^{i}$ represents a scalar field, and $\psi
^{i}$ represents a fermion in a 2-spinor representation. We make the
assumption that $G_{i}$ is much smaller than ${\mathcal{O}}(1)$ for all
fields, excluding the field responsible for SUSY breaking.
The presence of the SUSY breaking field may lead to the suppression of the
contribution proportional to $G_{\phi }$ during the inflaton decay due to
interference \cite{B13,B14}. In this section, we simplify our analysis by
assuming the minimal K\"{a}hler potential. Consequently, the K\"{a}hler
potential lacks non-renormalizable terms. Despite isolating the inflaton
field from other fields in the global SUSY Lagrangian, the presence of (SUGRA) corrections enables its decay. The coupling constants
are expanded at the vacuum state \cite{B15}.%
\begin{equation}
G_{\phi ijk}=-\frac{W_{\phi }}{W}\frac{W_{ijk}}{W}+\frac{W_{\phi ijk}}{W}%
\simeq K_{\phi }\frac{W_{ijk}}{W}+\frac{W_{\phi ijk}}{W}.
\end{equation}

In previous analysis \cite{B15}, the assumption was made that the
Vacuum Expectation Values (VEVs) are effectively negligible for all fields
except the inflaton. We employed the condition $G_{\phi }\ll \left\langle
\phi \right\rangle $ in the final equation. Notably, we observed that the
outcome remains invariant under K\"{a}hler transformation, and these
constants tend to zero in the global SUSY limit.
Subsequently, the decay rates are computed%
\begin{equation}
\Gamma _{3/2}\left( \phi \longrightarrow \psi ^{i}\psi ^{j}\varphi
^{k}\right) \simeq \frac{N_{f}}{1536\pi ^{3}}\left\vert Y_{ijk}\right\vert
^{2}\left( \frac{\left\langle \phi \right\rangle }{M_{p}}\right) ^{2}\frac{%
m_{\phi }^{3}}{M_{p}^{2}},
\end{equation}

Here, $N_{f}$ represents the count of final states, and the Yukawa coupling $%
Y_{ijk}$ is denoted as $W_{ijk}$. In this context, we have disregarded the
masses of the particles in the final state and employed $K=\varphi ^{\dagger
}\varphi $ for the inflaton. Additionally, it is assumed here that the particles $\psi ^{i}$ and $\psi ^{j}$ are non-identical . The decay rates of
the inflaton into scalar particles align with the previously mentioned
results. Indeed, considering the scalar potential, $V=e^{G}\left(
G^{j}G_{j}-3\right) ,$\ the estimated decay amplitude of $\phi ^{\ast
}\Longrightarrow \varphi ^{i}\varphi ^{j}\varphi ^{k}$\ is denoted as $V_{%
\bar{\phi}ijk}$. Given the sizable SUSY mass of the inflaton, this
amplitude is approximately proportional to $e^{G/2}G_{\phi ijk}$, multiplied
by the inflaton mass, $m_{\phi }\simeq e^{G/2}\left\vert G_{\bar{\phi}%
}^{\phi }\right\vert $. The inflaton's decay into a pair of gravitinos occurs at the rate $\Gamma$. The resulting gravitino abundance $Y_{3/2}$
is then calculated, with $Y_{3/2}$ equal to the ratio of the final gravitino
abundance to the entropy density $\frac{n_{3/2}}{s}$ \cite{B16,B17,B18}.%
\begin{equation}
Y_{3/2}\simeq 2\times 10^{-11}\left( \frac{10^{6}GeV}{T_{re}}\right) \left( 
\frac{m_{\phi }\times \left( \left\langle \phi \right\rangle \right) }{%
10^{27}GeV^{2}}\right) ^{2}.
\end{equation}%
\begin{figure}[tbp]
\centering
\includegraphics[width=18cm]{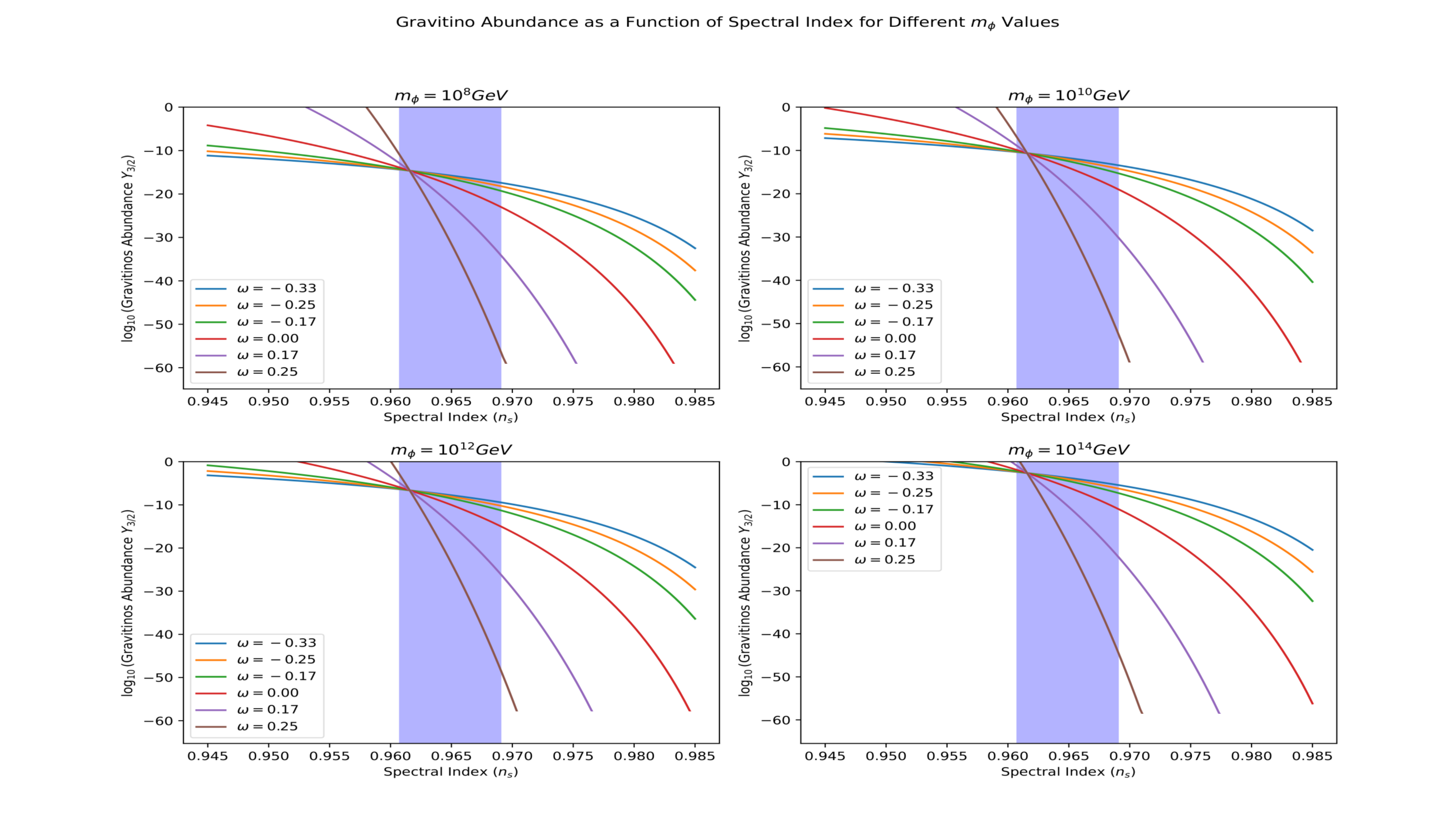}
\caption{{} The gravitinos abundance $Y_{3/2}$\ as functions of the spectral
index $n_{s}$ for different values of the equation of state $\protect\omega %
. $}
\label{fig:1}
\end{figure}

The plot in Fig. \ref{fig:1} displays the gravitino abundance $Y_{3/2}$ as a
function of the spectral index $n_{s}$ for different values of the inflaton
mass $m_{\phi }$. Each subplot corresponds to a specific $m_{\phi }${}
values, ranging from $10^{8}GeV$ to $10^{14}GeV$\. Different curves
representing various equation of state parameter values during reheating $%
\omega $, with a shaded region highlighting the Planck bound
on $n_{s}=0.9649\pm 0.0042$\ according to recent observations \cite{B19}.

From the given formalism, the abundance of\ gravitinos produced during the
reheating phase, denoted as $Y_{3/2}$ is significantly dependent on the
reheating temperature $T_{re}$,which is affected by $\omega $. The curves of
the gravitinos abundance tend towards the central value where all the
lines with the different equation of state values converge. This central
value of the abundance curves corresponds to the highest reheating
temperature, and all the lines coincide within the observed
bound on the spectral index $n_{s}$. Higher $m_{\phi }$ values generally
lead to increased gravitino production when considering the maximum
reheating temperature, while different $\omega $ values impact the reheating
temperature and thus the gravitino abundance, causing $Y_{3/2}$ to decrease  away from the observed bound of $n_{s}$. 

In fact, lower equation of state values
tend to fall outside the $n_{s}$ bound faster than higher values of $\omega $.
Consequently, this plot underscores the need to reconcile reheating
scenarios with observational constraints on gravitino abundance.

\section{Gravitational Waves Background}

\label{sec4}

\subsection{Scalar induced gravitational waves}

Detecting the background of primordial gravitational waves would provide
strong evidence for the inflation paradigm and offer insights into the
fundamental physics of the early Universe \cite{Ade:2015lrj}. Recent studies have focused on
Primordial Gravitational Waves (PGW) from the period immediately following
inflation \cite{B20,B21,B22,K1,K2,K3,K4,K5,K6,K7,K8,K8-1}. Within the inflationary scenario, the presence of a kinetic epoch
preceding inflation leads to a distinct blue tilt in the spectra of
primordial gravitational waves at higher frequencies \cite{B23}. Gravitational waves
are identified as the transverse-traceless component of metric
perturbations. According to linear perturbation theory, scalar, vector, and
tensor modes do not interact. Now, let us calculate the gravitational waves
generated by second-order gravitational interactions resulting from
first-order curvature perturbations \cite{C01,C02,C03,C04}. Using the
Newtonian gauge, the perturbed metric is expressed as \cite{C1}:%
\begin{equation}
ds^{2}=a^{2}\left( \tau \right) \left[ -\left( 1+2\Phi \right) d\tau
^{2}+\left(\delta_{i j} \left( 1-2\Phi \right) +\frac{h_{ij}}{2}\right) dx^{i}dx^{j}%
\right] ,
\end{equation}

where $\tau$ is the conformal time, $\Phi $ represents the first-order Bardeen gravitational potential,
and $h_{ij}$ denotes the second-order tensor perturbation. We can
reformulate the spectral abundance of gravitational waves, defined as the
energy density of gravitational waves per logarithmic comoving scale, as 
\cite{C2,C3},

\begin{equation}
\Omega _{GW}(k)=\frac{k^{2}}{12H_{0}}{\mathcal{T}}(\mathbf{\tau }_{0},%
\mathbf{k}){\mathcal{P}}_{t}(k).
\end{equation}

The time evolution of a gravitational wave field, denoted as $h_{\mathbf{k}%
}(\tau _{i})$ at an initial conformal time $\tau _{i}$ and characterized by
its tensor spectrum, can be determined by computing the GW transfer function 
${\mathcal{T}}\mathbf{(\tau ,k)=}h_{\mathbf{k}}(\tau )/h_{\mathbf{k}}(\tau
_{i})$ \cite{C4}. Here, $h_{\mathbf{k}}(\tau )$ is evaluated at a conformal
time $\eta \gg \eta _{i}$ \cite{C5,C6}. The current conformal time is denoted by $\mathbf{\tau }_{0}$, and the Hubble constant is represented by $H_{0}$ \cite{C7}. Our focus is  on the spectral energy density
parameter\ $\Omega _{GW}(k)$\ at Pulsar Timing Array (PTA) scales, where $%
f\sim {\mathcal{O}}(10-9)Hz$ corresponds to wavenumbers $k\sim {\mathcal{O}}%
(10^{6})Mpc^{-1}$\, which are significantly larger than the wavenumber linked
to a mode crossing the horizon at matter-radiation equality. Stated
differently, the modes observed at PTA scales crossed the horizon deep
within the radiation era, well before matter-radiation equality. In the
regime where $k\gg k_{eq},$\ the gravitational wave spectral energy density
associated with PTA signals can be expressed as \cite{C8,C9,C10,C11,C12,C13}%
\begin{equation}
\Omega _{GW}(f)=\frac{2\pi ^{2}f_{yr}^{2}}{3H_{0}^{2}}A^{2}\left( \frac{f}{%
f_{yr}}\right) ^{\alpha }. \label{33}
\end{equation}

The NANOGrav measurements \cite{C14,C15}, estimate ranges of  several parameters which are given as follows, $\alpha =\left( 5-\gamma \right)$ with $\gamma=13/3$, $f_{yr}$ is best estimated to $f_{yr}\simeq 3.1\times 10^{-8}$
and $H_{0}$\ is given by $H_{0}\equiv 100h~km/s/Mpc$. The relationship
linking the amplitude of the Pulsar
Timing Array (PTA) signal, with the
cosmological parameters were found to be \cite{C16,C17}%
\begin{equation}
A=\sqrt{\frac{45\Omega _{m}^{2}A_{s}}{32\pi ^{2}(\eta _{0}k_{eq})^{2}}}c%
\frac{f_{yr}}{\eta _{0}}\left( \frac{f_{yr}^{-1}}{f_{\star }}\right) ^{\frac{%
n_{T}}{2}}\sqrt{r}.
\end{equation}

The dependence on $n_{T}$ in this context arises from the substantial "lever arm" between the Cosmic Microwave Background (CMB) pivot frequency, where $A_{s}$ is constrained, and the frequency of the Pulsar Timing Array (PTA) signal $%
\left[ 1yr^{-1}\right] $.

\begin{table*}[tbp]
\centering%
\begin{tabular}{c|c|c|c||c|c}
\hline
$g$ & $\sigma $ & $r$ & $A$ & $f(Hz)$ & $h^{2}\Omega _{gw}(f)$ \\ 
\hline\hline
$3\times 10^{-3}$ & $2\times 10^{-5}$ & $0.001$ & $1.44\times 10^{-15}$ & $%
4\times 10^{-8}$ & $2.91\times 10^{-9}$ \\ \hline
$2\times 10^{-3}$ & $3\times 10^{-6}$ & $0.009$ & $4.33\times 10^{-15}$ & $%
2\times 10^{-8}$ & $1.85\times 10^{-8}$ \\ \hline
$4\times 10^{-3}$ & $7\times 10^{-6}$ & $0.026$ & $9.38\times 10^{-15}$ & $%
1\times 10^{-8}$ & $3.78\times 10^{-8}$ \\ \hline
$5\times 10^{-3}$ & $9\times 10^{-6}$ & $0.039$ & $1.03\times 10^{-14}$ & $%
8\times 10^{-7}$ & $5.07\times 10^{-7}$ \\ \hline
$4\times 10^{-3}$ & $5\times 10^{-6}$ & $0.052$ & $1.04\times 10^{-14}$ & $%
6\times 10^{-7}$ & $5.86\times 10^{-7}$ \\ \hline
$5\times 10^{-3}$ & $5\times 10^{-6}$ & $0.128$ & $1.63\times 10^{-14}$ & $%
4\times 10^{-7}$ & $1.17\times 10^{-6}$ \\ \hline
\end{tabular}%
\caption{Testing the density of gravitational waves predicted by PTA as
functions of several parameters related to the D-term hybrid model. Here, we
have incorporated the best-fit values for the cosmological parameters
according to the Planck results \protect\cite{B19}, $\Omega _{m}=0.315,$\ $%
\protect\eta _{0}\simeq 1.38,$\ $A_{s}=2,1\times 10^{-9},k_{eq}\simeq
0.01,f_{\star }\approx 7.7\times 10^{-17}Hz,$\ $f_{yr}\simeq 3.1\times
10^{-8}Hz$ and $H_{0}\equiv 100~h~km/s/Mpc.$\ \ \ \ \ \ \ \ \ }
\label{table:1}
\end{table*}

Table \ref{table:1} presents the NanoGrav proposed model for the density $%
h^{2}\Omega _{gw}(f)$ as a function of frequency $f,$ scalar-to-tensor ratio 
$r$, and the amplitude of the PTA signal related to the potential parameters
of the D-term hybrid model. The model for $\Omega _{gw}(f)$ depends on
various cosmological parameters and constants defined earlier. The
analytical expression for, $h^{2}\Omega _{gw}(f)$ is derived from a recent
theoretical model for primordial gravitational waves. $h^{2}\Omega _{gw}(f)$
represents the fractional energy density of gravitational waves in the
universe, indicating the energy carried by the gravitational wave background
relative to the critical energy density needed for a flat universe. By
examining the table \ref{table:1}, one can identify the parameters that predict higher or
lower densities of gravitational waves $h^{2}\Omega _{gw}(f)$ in line with
NANOGrav predictions \cite{C6-1,C6-2,C6-3,C6-4}. The amplitude of PTA scales $A\lesssim 10^{-20}$
provides accurate predictions for $h^{2}\Omega _{gw}(f)$ values.
Furthermore, the Table \ref{table:1} offers insights into the behavior of gravitational
wave background (GWB) density concerning frequency, scalar-to-tensor ratio,
and the D-term model parameters, demonstrating good consistency with
predicted bounds on density and frequency when fine-tuning the inflationary
parameters associated with PTA signals.

\subsection{Primordial Gravitational Waves}

For primordial gravitational waves in a spatially flat
FLRW background, the metric element can be expressed as follows:%
\begin{equation}
ds^{2}=a^{2}(\tau)\left[-d\tau^2+\left(\delta _{i j}+h_{i j }\right)dx^{i }dx^{j}\right].
\end{equation}
The perturbation $h_{i j }$ satisfies the transverse-traceless (TT) conditions: $h^{i}_{~i}=0$ and $\partial^{i }h_{i j }=0$. The tensor perturbation $h_{ij }\left( \tau,\vec{x}\right) $\ can be decomposed into its Fourier modes, which are associated with two polarization tensors, satisfying the equation of motion
\begin{equation}
h_{\mathbf{k}}^{\lambda \prime \prime }(\tau )+2{\mathcal{H}}h_{\mathbf{k}%
}^{\lambda \prime }(\tau )+k^{2}h_{\mathbf{k}}^{\lambda }(\tau )=0.
\end{equation}%
Here $\left( ^{\prime }\right) $ denotes the derivative with respect to
conformal time $\tau$, where $d\tau =$\ $\frac{dt}{a}$ and ${\mathcal{H}}=%
\frac{a^{\prime }}{a}$. The normalized gravitational wave energy density
spectrum is defined as the energy density per logarithmic frequency interval
\begin{equation}
\Omega _{gw}(k)=\frac{1}{\rho _{c}}\frac{d\rho _{gw}}{d\ln k},
\end{equation}%
here $\rho _{c}$ represents the energy density. Additionally,%
\begin{equation}
\Omega _{gw,0}(k)=\frac{1}{12}\left( \frac{k^{2}}{a_{0}^{2}H_{0}^{2}}\right) 
{\mathcal{P}}_{h}(k),
\end{equation}%
knowing that ${\mathcal{P}}_{h}(k)\equiv \frac{k^{3}}{\pi ^{2}}\sum_{\lambda
}\left\vert h_{\mathbf{k}}^{\lambda }\right\vert ^{2}.$ By considering the
scale at which horizon re-entry occurs $\left( k=a_{hc}H_{hc}\right) \ $and
analyzing the horizon re-entry scale alongside the Hubble parameter across
different epochs, we can derive the present-day primordial gravitational
wave spectrum for mode re-entry during the matter-dominated (\textit{M}),
radiation-dominated (\textit{R}), and kinetic (\textit{K}) eras,
respectively, as follows:\ 
\begin{eqnarray}
\Omega _{gw,0}^{\left( \mathit{M}\right) } &=&\frac{1}{24}\Omega _{m,0}^{2}%
\frac{a_{0}^{2}H_{0}^{2}}{k^{2}}{\mathcal{P}}_{t}~~~~~~\left( k_{0}<k\leq
k_{eq}\right) , \\
\Omega _{gw,0}^{\left( \mathit{R}\right) } &=&\frac{1}{24}\Omega
_{r,0}^{2}\left( \frac{g_{\ast }}{g_{\ast 0}}\right) \left( \frac{g_{\ast s}%
}{g_{\ast s0}}\right) {\mathcal{P}}_{t}~~~~~~\left( k_{eq}<k\leq
k_{r}\right) , \\
\Omega _{gw,0}^{\left( \mathit{K}\right) } &=&\Omega _{gw,0}^{\left( \mathit{%
R}\right) }\left( \frac{k}{k_{r}}\right) ~~~~~\left( k_{r}<k\leq k_{\max
}\right) ,
\end{eqnarray}

\begin{figure}[tbp]
\centering
\includegraphics[width=18cm]{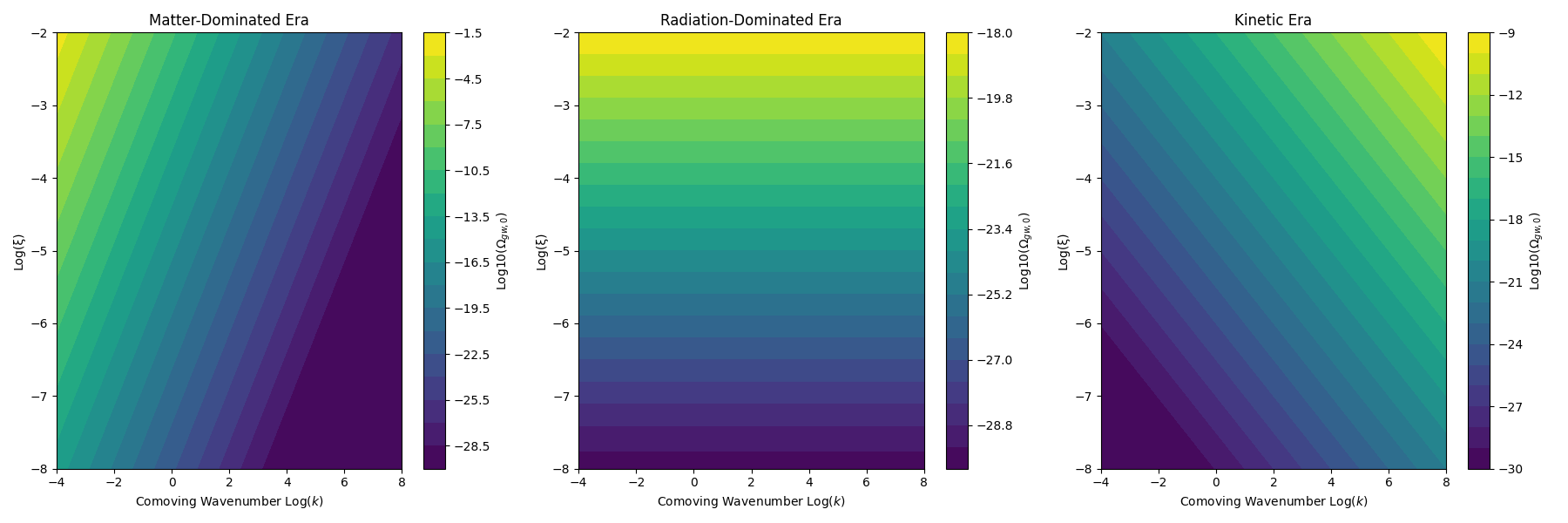}
\caption{{}D-term $\protect\xi $ parameter as a function of the comoving
wavenumber $k$ across the primordial gravitational wave spectrum $\Omega
_{gw,0}$.}
\label{fig:2}
\end{figure}

Fig. \ref{fig:2} provides a detailed visualization of the primordial tensor
power spectrum and the corresponding energy density parameters $\Omega
_{gw,0}$\ across different cosmological eras: matter-dominated,
radiation-dominated, and kinetic eras. In the matter-dominated era, $\Omega
_{gw,0}$\ decreases with increasing comoving wavenumber $k$, indicating
larger scales are more influenced by primordial tensor perturbations. During
the radiation-dominated era, the energy density of gravitational waves is only affected directly by the D-term hybrid parameter $\protect\xi $, while the scale $k$ has a constant effect on $\Omega_{gw,0}$. In the kinetic era, the energy density of gravitational waves is increasing with respect to the comoving scale to reach its maximum values at superhorizon scales. This
comprehensive depiction illustrates how primordial gravitational waves
evolve through different epochs, influenced by the universe's thermal
history and expansion dynamics, and underscores the significance of these
factors on the energy densities observed.\ 

\section{Constraints on Gravitational Waves from Cosmic Strings}

\label{sec5}

Cosmic strings form at the end of inflation, impacting the
anisotropies observed in CMB and contributing to
the creation of stochastic gravitational waves \cite{J1}. The dimensionless string
tension, $G\mu _{cs},$ is key to understanding these phenomena, where $G=%
\frac{1}{8\pi M_{p}^{2}}=6.7\times 10^{-39}GeV^{-2},$\ and $\mu _{cs}$ is
the string's mass per unit length. Current CMB constraints place limits on
this tension as $G\mu _{cs}\lesssim 1.3\times 10^{-7}$ \cite{J1}. The SGWB  arises from a mix of sources including inflation,
cosmic strings, and phase transitions \cite{J2}. Specifically, inflationary tensor
perturbations re-entering the horizon generate an SGWB \cite{J3,J4,J5,J6},
leaving a unique imprint on the CMB B-mode polarization. The amplitude and
scale dependence of this background are described by the tensor-to-scalar
ratio $r$ and the tensor spectral index\ $n_{T},$\ adhering to the
inflationary consistency relation $r=-8n_{T}$\ \cite{J7}.\ Given that $r\geq
0,$\ this implies $n_{T}\leq 0$\ indicating a red spectrum \cite{J8}. With
current limits on the tensor-to-scalar ratio, the amplitude of the
inflationary SGWB at pulsar timing array (PTA) and interferometer scales
remains too low for detection by these instruments, necessitating a
primordial tensor power spectrum with a strong blue tilt $\left( n_{T}\geq
0\right) $\ for detection \cite{J9,J10}.

The detection of gravitational waves from cosmic strings is primarily
influenced by two key scales: the energy scale of inflation $\Lambda _{\inf
} $\ and the scale at which cosmic strings generate the GW spectrum $\Lambda
_{cs}\equiv \sqrt{G\mu _{cs}}$. The amplitude of the tensor mode anisotropy
in the cosmic microwave background fixes the energy scale of inflation to
approximately $\Lambda _{\inf }\sim V^{1/4}\sim 3.3\times 10^{16}r^{1/4}$\ 
\cite{J11}. By applying the Planck $2\sigma $ bounds on the tensor-to-scalar
ratio $r$, we derive an upper limit on the inflation scale, $\Lambda _{\inf
}<1.6\times 10^{16}GeV$\ \cite{J12}. In our model, cosmic strings form
post-inflation, indicating $\Lambda _{\inf }>\Lambda _{cs},$\ which results
in a SGWB generated from
undiluted strings. The SGWB from metastable cosmic string networks is
expressed relative to the critical density as described in the folowing form 
\cite{J13},%
\begin{eqnarray}
\Omega _{GW}\left( f\right) &=&\frac{8\pi G}{3H_{0}^{2}}f\left( G\mu
_{cs}\right) ^{2}\Sigma _{n=1}^{\infty }C_{n}\left( f\right) P_{n},  \notag
\\
&=&\frac{8\pi G}{3H_{0}^{2}}f{\mathcal{G}}_{cs}(C_{n},P_{n}). \label{40}
\end{eqnarray}%
Here, ${\mathcal{G}}_{cs}(C_{n},P_{n})$\ represent the term of cosmic
strings contribution to GWs production, which contain the power spectrum $%
P_{n}\simeq \frac{50}{\zeta \left( 4/3\right) n^{4/3}}$ that represent the
gravitational waves (GWs) emitted by the \textit{n-th} harmonic of a cosmic
string loop, and $C_{n}$ which denotes the number of loops emitting GWs
observed at a specific frequency $f$\ . The number of loops emitting GWs,
observed at a given frequency $f$ is defined as \cite{J13},%
\begin{equation}
C_{n}\left( f\right) =\frac{2n}{f^{2}}\int_{z_{\min }}^{z_{\max }}\frac{dz}{%
H(z)\left( 1+z\right) ^{6}}{\mathcal{N}}(\mathit{l},t).
\end{equation}

The integration range spans the lifetime of the cosmic string network,
starting from its formation at $z_{\max }\simeq \frac{T_{R}}{2.7K},$\ with $%
T_{R}$ being approximately $10^{9}GeV$ to its decay at $z_{\min }=\left( 
\frac{70}{H_{0}}\right) ^{1/2}\left( \frac{\Gamma\left( G\mu _{cs}\right)
^{2}}{2\pi \times 6.7\times 10^{-39}}\exp \left( -\pi \kappa _{cs}\right)
\right) ^{1/4}$\ \cite{J14}, where $\Gamma \simeq 50$ is a numerical factor specifying the cosmic strings decay rate. Here, ${\mathcal{N}}(\mathit{l},t)$ represents
the number density of CS loops of length $\mathit{l}=\frac{2n}{\left(
1+z\right) f}.$ The loop density is defined by considering their formation
and decay across different epochs. For the region of interest, the dominant
contribution arises from the loops generated during the radiation-dominated
era which is given by \cite{J14},%
\begin{equation}
{\mathcal{N}}_{r}(\mathit{l},t)=\frac{0.18}{t^{2/3}\left( \mathit{l}+\Gamma
G\mu _{cs}t\right) ^{5/2}}.
\end{equation}
The cosmological time and the Hubble rate with the current values of matter,
radiation and dark energy densities, are respectively expressed as a function of the redshift $z$ as $t(z)=\int_{z_{\min }}^{z_{\max }}\frac{dz}{%
H(z)\left( 1+z\right) },$ $H(z)=H_{0}\sqrt{\Omega _{\Lambda }+\Omega
_{m}\left( 1+z\right) ^{3}+\Omega _{r}\left( 1+z\right) ^{4}}.$ \ \ 

\begin{figure}[tbp]
\centering
\includegraphics[width=16cm]{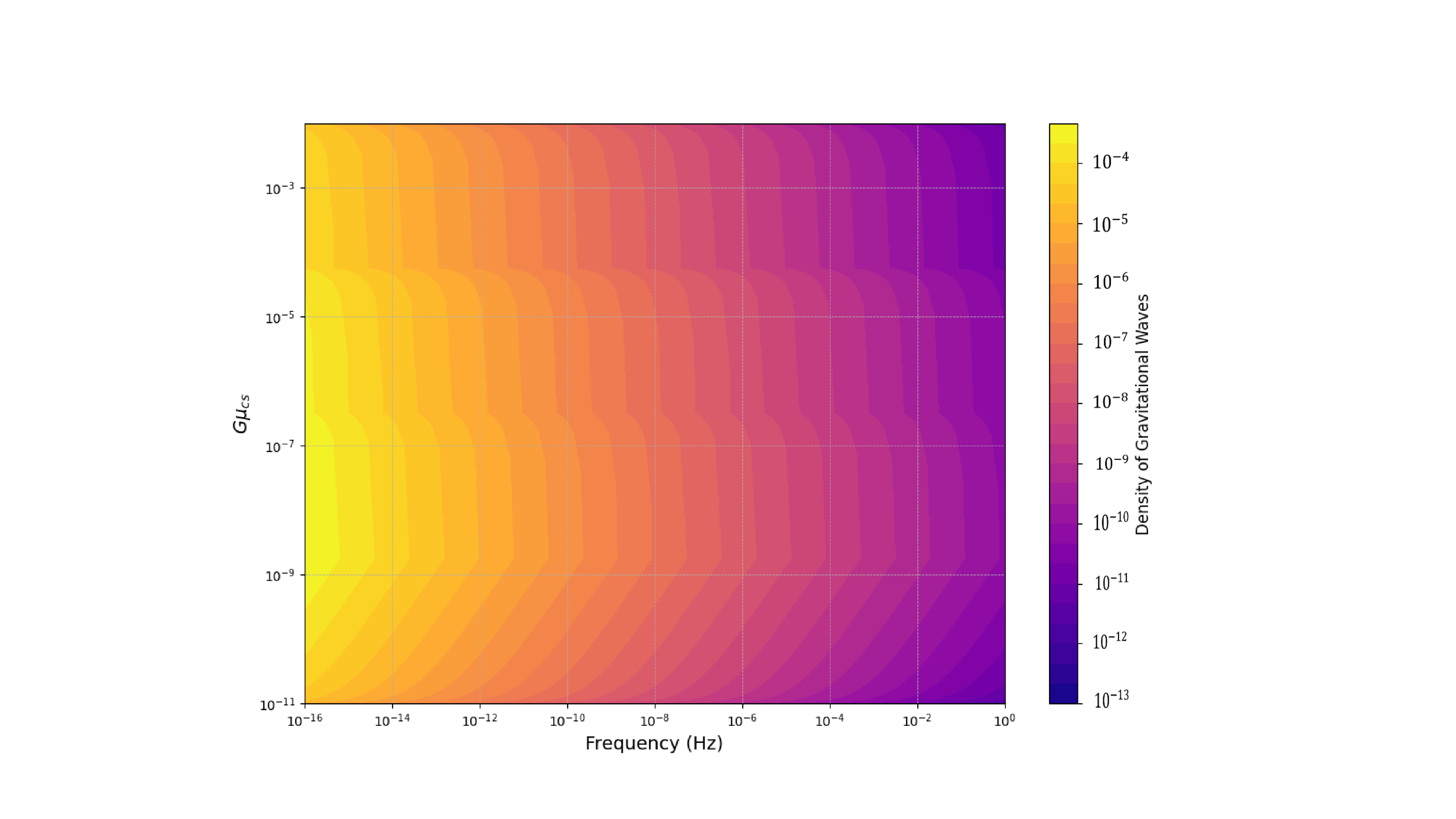}
\caption{{}Gravitational waves density $\Omega _{GW}$\ as a function of the
frequency $f$ and the cosmic string parameter $G\protect\mu _{cs}.$}
\label{fig:3}
\end{figure}

Fig. \ref{fig:3} illustrates the density of gravitational waves $\Omega
_{GW} $\ as a function of frequency $f$ and the cosmic string parameter $%
G\mu _{cs} $. The contour plot uses a logarithmic scale for both the
frequency and $G\mu _{cs}$, highlighting how gravitational wave densities
vary across different scales and parameters. The gravitational wave energy
density is computed by integrating over the redshift $z$, taking into
account the contributions from matter, radiation, and dark energy to the
Hubble parameter $H(z)$ The function ${\mathcal{N}}_{r}(\mathit{l},t)$
reflects the number density of gravitational waves, which is influenced by
the redshift and other cosmological parameters.\ $C_{n}$\ and $P_{n}$\ are
functions of $n $\ , frequency $f$, and the cosmic string parameter ${%
\mathcal{G}}_{cs},$\ representing cosmic strings contribution to the power
spectrum of the gravitational waves.\ The resulting plot shows that at
higher frequencies, the gravitational wave density varies more significantly
with changes in $G\mu _{cs},$ indicating a strong dependence on the string
tension parameter. This comprehensive depiction underscores the intricate
relationship between cosmic string dynamics and gravitational wave
emissions, offering insights into how different frequencies and cosmic
string tensions contribute to the observable gravitational wave background.
The use of a logarithmic scale for both axes ensures that a wide range of
values is represented, making it easier to visualize the detailed structure
and variation of $\Omega _{GW}$\ across different cosmological scenarios.\ \
\ \ \ \ \ \ \ \ 

\begin{figure}[tbp]
\centering
\includegraphics[width=10cm]{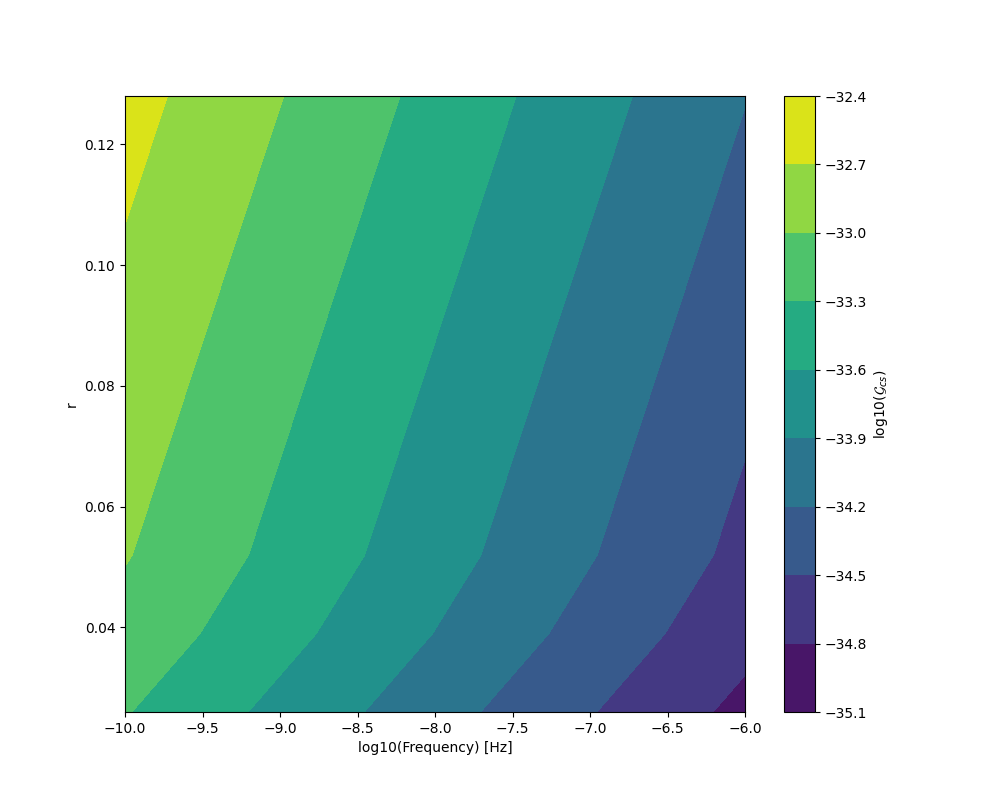}
\caption{{}Cosmic strings factor ${\mathcal{G}}_{cs}$\ as functions of the
frequency $f$ and the tensor-to-scalar ratio $r$. }
\label{fig:4}
\end{figure}

Fig. \ref{fig:4} presents the cosmic strings factor ${\mathcal{G}}_{cs}$\ as
a function of frequency and the tensor-to-scalar ratio $r$ combining
equations (\ref{33}) and (\ref{40}). The contour plot
uses a logarithmic scale for frequency, ranging from $10^{-10}Hz$ to $%
10^{-6}Hz$\ , and spans multiple values of $r=\left[ 0.026,0.128\right] $,
reflecting different possible strengths of primordial tensor perturbations.
The frequency dependence of ${\mathcal{G}}_{cs}$ incorporates a power-law
behavior with an exponent $\alpha $, representing the spectral shape of the
cosmic string signal. The resulting plot visually demonstrates that ${%
\mathcal{G}}_{cs}$\ increases with higher values of $r$ and decreases with
increasing frequency. The contour levels indicate the logarithmic values of,
with color gradients representing the intensity. This visualization helps to
elucidate the relationship between cosmic string dynamics and gravitational
wave signals across a range of frequencies and tensor-to-scalar ratios.
Specifically, it shows that stronger primordial tensor perturbations lead to
higher ${\mathcal{G}}_{cs}$\ values, particularly at lower frequencies,
which is crucial for understanding the potential observational signatures of
cosmic strings in gravitational wave experiments. This comprehensive
depiction provides insights into the scale and strength of cosmic string
contributions to the gravitational wave background, highlighting key
dependencies and aiding in the interpretation of potential observational
data in the context of early universe cosmology.\ 

\section{Conclusion}

\label{sec6}

In this paper, we investigated the framework of supergravity model,
employing the minimal K\"{a}hler potential. Based on previous studies, we
recall the formalism that demonstrated how the D-term potential can resolve
the $\eta $-problem inherent in F-term models, offering a viable path to
successful inflation. We examined the scalar potential that includes both
F-term and D-term contributions which demonstrated a stability conditions
for the inflaton field $S$. The effective potential during inflation was
approximated, highlighting the importance of the gauge coupling constant $g$
and the Fayet-Iliopoulos term $\xi $. The slow-roll parameters and the
resulting e-folding number were calculated to test the inflationary
dynamics. Furthermore, we explored reheating dynamics and gravitino
production, emphasizing the interplay between reheating temperature $T_{re}$%
, spectral index $n_{s}$, and gravitino abundance $Y_{3/2}$. Our analysis
indicated that gravitino production is sensitive to the equation of state
during reheating, impacting the reheating temperature and the subsequent
dark matter relic density. Moreover, we examined the density of
gravitational waves $h^{2}\Omega _{gw}(f)${} predicted by Pulsar Timing
Array experiments, as a function of several parameters within the D-term
hybrid model. Our results align well with PTA constraints, particularly in
terms of the amplitude of the scalar-to-tensor ratio $r$ and its relation to
the fractional energy density of gravitational waves. These findings provide
further constraints on inflationary parameters and the spectral energy
density of primordial gravitational waves. The subsequent results offers
insights into the primordial gravitational wave spectrum across different
cosmological eras, we found that the energy density $\Omega _{gw,0}$\
evolves distinctly through matter, radiation, and kinetic eras, heavily
influenced by the D-term parameter $\xi.$\ This highlights the significance
of primordial gravitational waves in the evolution of the early Universe and
the role of inflationary models in shaping these waves. The relationship
between the cosmic string tension $G\mu _{cs}$\ and gravitational wave
density, reveals that cosmic strings can significantly affect the SGWB at
higher frequencies. The strong dependence of gravitational wave density on
the string tension parameter emphasizes the potential observational
signatures of cosmic strings, offering a window into the dynamics of the
early Universe and cosmic string properties.

Our findings provide insights into the role of supergravity-based inflation
models in the generation and evolution of gravitinos, cosmic strings, and
gravitational waves. The SGWB produced by cosmic strings and its sensitivity
to parameters such as the string tension $G\mu _{cs}$\ and reheating
dynamics offer valuable constraints on early Universe dynamics. Furthermore,
our results demonstrate the relevance of these models in current and future
gravitational wave observations, such as those from PTA experiments, and
their potential to uncover new physics related to dark matter and the cosmic
microwave background.

\section*{Acknowledgments}

G.O. acknowledges the financial support of Fondecyt Grant 1220065.


\end{document}